\documentclass[conference]{IEEEtran}
\usepackage{epsf}
\usepackage{graphicx}
\usepackage{graphics}
\usepackage{amsmath,bm}
\usepackage{amssymb}
\usepackage{epsfig,latexsym,amsmath,epsf,amssymb,amsfonts}
\usepackage{verbatim}
\usepackage{placeins}
\usepackage[hang]{subfigure}
\usepackage{epstopdf}

\usepackage{amsthm}

\newcommand{\ignore}[1]{}

\usepackage{url}

\begin{document}
%\title{Stable Throughput Region of Cooperative Communication in Cognitive Radio Networks with Finite Relaying Buffer Capacity}
%Wireless Intelligent Networks Center (WINC), School of Communication and Information Technology,\\
%Nile University, Giza 12677, Egypt.\\
\title{On the Stable Throughput of Cooperative Cognitive Radio Networks with Finite Relaying Buffer}
\author{\large Adel M. Elmahdy$^{\dag}$, Amr El-Keyi$^{\dag}$, Tamer ElBatt$^{\dag\S}$ and Karim G. Seddik$^*$\\ [.1in]
\normalsize
\begin{tabular}{c}
$^{\dag}$Wireless Intelligent Networks Center (WINC), Nile University, Giza, Egypt.\\
$^{\S}$EECE Dept., Faculty of Engineering, Cairo University, Giza, Egypt.\\
$^*$Electronics Engineering Department, American University in Cairo, AUC Avenue, New Cairo 11835, Egypt.\\
Email: adel.elmahdy@ieee.org, aelkeyi@nileuniversity.edu.eg, telbatt@ieee.org, kseddik@aucegypt.edu
\end{tabular}
}
\maketitle

%%%%%%%%%%%%%%%%%%%%%%%%%%%%%%%%%%%%%%%%%%%%%%%%%%%%%%%%%%%%%%%%%%%%%%%%%%%%%%%%%%%%%%%%%%%%%%%%%%%%%%%%%%%%%%%%%%%%%%%%%%%
\begin{abstract}
In this paper, we study the problem of cooperative communications
in cognitive radio systems where the secondary user has limited
relaying room for the overheard primary packets. More
specifically, we characterize the stable throughput region of a
cognitive radio network with a finite relaying buffer at the
secondary user. Towards this objective, we formulate a constrained
optimization problem for maximizing the secondary user throughput 
while guaranteeing the stability of the primary user
queue. We consider a general cooperation policy where the packet
admission and queue selection probabilities, at the secondary
user, are both dependent on the state (length) of the finite
relaying buffer. Despite the sheer complexity of the optimization
problem, attributed to its non-convexity, we transform it to a 
linear program. Our numerical
results reveal a number of valuable insights, e.g., it is always
mutually beneficial to cooperate in delivering the primary packets
in terms of expanding the stable throughput region. In addition,
the stable throughput region of the system, compared to the case of 
infinite relaying queue capacity, marginally shrinks for limited relaying queue capacity.

%the amount of enhancement depends on the finite length of the
%relaying queue of the secondary user.
%We investigate the performance of cooperative communication in
%cognitive radio systems. Specifically, we focus on identifying the
%stable throughput region of a cognitive radio network with a
%finite relaying buffer at the secondary user.  In the proposed
%cooperation policy, the packet relaying probabilities are
%functions of the number of packets in the finite relaying buffer.
%The problem of finding the maximum stable throughput of the
%secondary user is formulated as a constrained non-convex
%optimization problem. Yet, we transform it into a linear program
%via standard techniques. Simulation results show the effect of the
%finite queue length on the stable throughput region of the system.
\end{abstract}


%%%%%%%%%%%%%%%%%%%%%%%%%%%%%%%%%%%%%%%%%%%%%%%%%%%%%%%%%%%%%%%%%%%%%%%%%%%%%%%%%%%%%%%%%%%%%%%%%%%%%%%%%%%%%%%%%%%%%%%%%%%
\section{Introduction}

%\vspace{-10pt}
\makeatletter{\renewcommand*{\@makefnmark}{}
\footnotetext{\hrule \vspace{0.05in} 
This work was made possible by grants number NPRP 4-1034-2-385 and 
NPRP 5-782-2-322 from the Qatar National Research Fund (a member of
Qatar Foundation). The statements made herein are solely the responsibility
of the authors.
}\makeatother}

The concept of cognitive radios was motivated by the
problem of spectrum scarcity, as well as the inefficient
utilization of the licensed spectrum  \cite{Mitola}, \cite{H_CR}.
The fundamental purpose of deploying cognitive radio networks is
to enhance the spectrum utilization by exploiting spectral holes
where a portion of the licensed spectrum is unutilized for a
period of time. Accordingly, the cognitive radio technology proposes 
the coexistence of unlicensed secondary users (SUs) with the licensed
primary users (PUs) on the same frequency band in such a way that
a minimum quality of service (QoS) is guaranteed for the PUs. For
example, in overlay cognitive radio networks, the SU seizes the 
opportunity to transmit its packets when it senses a spectrum hole, 
that is, the PU is idle.

In wireless networks, the notion of cooperative communications has
attracted considerable attention in recent years. The  basic idea 
relies on the broadcast nature of the wireless
channel where intermediate nodes between the transmitter and 
receiver can be thought of as potential relay nodes. If a receiver
fails to decode a packet, there is a good chance that one of the
intermediate nodes has received this packet and, hence, can aid
in delivering it to the receiver. Therefore, cooperative
communications can play a major role in enhancing the performance
of wireless networks. Cooperative communications has been explored 
within the framework of cognitive radio networks whereby the SU 
helps the PU in delivering its packets to the destination \cite{Tse,Kramer}.
Intuitively, both the PU and the SU benefit from such cooperation.
The PU reliably delivers its own packets while the SU witnesses an 
increase in the number of time slots available for transmission.

A considerable part of the wireless literature was dedicated to the
concept of cooperative communications from the perspective of the
physical layer, e.g., \cite{Tse}, \cite{Kramer}, \cite{Gamal}. Others
studied it from the perspective of the MAC layer, e.g., \cite{Sadek},
\cite{Krikidis}. Our prime focus in this paper is on cooperative
relaying in cognitive radio networks at the packet level.
Essentially, we extend the work presented in
\cite{R_CoopAcc}, \cite{Ashour_Conf} and \cite{Ashour_Journal}. In
\cite{R_CoopAcc}, a new protocol-level full cooperation in a wireless
multiple-access system is proposed for a system composed of $N$
users in which each node is a source, and at the same
time a potential relay. 
%Unlike this work, the secondary user 
%always helps the primary, gives strict priority to the relay 
%queue and the secondary relay queue is assumed to be infinite.
In \cite{Ashour_Conf,Ashour_Journal}, the authors propose a 
cooperative strategy with probabilistic relaying. In this
strategy, it is considered that the SU has two
\emph{infinite-length} queues; one is for its own packets and the
other is for relaying the PU packets. Upon overhearing a PU transmission, 
the SU enqueues the PU packet, if not correctly decoded by the destination, 
with probability (w.p.) $a$. Conversely, when the PU
is sensed idle, the SU serves its own data queue w.p. $b$ or the
relaying queue w.p. $1-b$.

Our main contribution in this paper is to characterize the stable throughput 
region of the system in \cite{Ashour_Journal} when the relaying buffer at the SU
has finite capacity. This, in turn, renders the problem more 
challenging due to the added complexity of modelling the state 
of the SU relay queue and solving the associated optimization problem. 
The motivation behind our work is to 
investigate the case when the SU helps the PU in
delivering its packets, yet, with limited resources, as opposed to
unlimited resources in \cite{R_CoopAcc}, \cite{Ashour_Journal}. It
can be contemplated that the proposed system model constitutes an important 
step towards real systems. Furthermore, we extend the system degrees of
freedom whereby the probability of the SU enqueueing an overheard 
PU packet is dependent on the state of the relaying queue, i.e., 
relaying queue length at a given time slot. In addition, 
the SU queue selection probabilities, between the relay queue and 
its own packets queue, is also assumed to be state-dependent.
%The stable throughput region is characterized via maximizing the SU
%throughput subject to constraints on the stability of all queues in the
%system for a given PU throughput. 
This problem formulation yields
a non-convex optimization problem. Nevertheless, we exploit the
structure of the problem that enables us to transform it into a
linear program that can be efficiently solved. We study
the effect of the finite size of the relaying queue on the stable
throughput region of the system. The numerical results show that 
the stable throughput region, compared to the case of
infinite relaying buffer size, marginally shrinks for
relatively small relaying queue lengths.

The rest of this paper is organized as follows. In Section II,
the system model is described. Section III demonstrates the proposed
cooperation policy. The characterization of the stable
throughput region of the system as an optimization problem is investigated
in Section IV. Numerical results are presented in Section V.
Finally, Section VI concludes the work and points out potential 
directions for future research.

%%%%%%%%%%%%%%%%%%%%%%%%%%%%%%%%%%%%%%%%%%%%%%%%%%%%%%%%%%%%%%%%%%%%%%%%%%%%%%%%%%%%%%%%%%%%%%%%%%%%%%%%%%%%%%%%%%%%%%%%%%%
\section{System Model}
\begin{figure}
\centering
\includegraphics[width=1\linewidth]{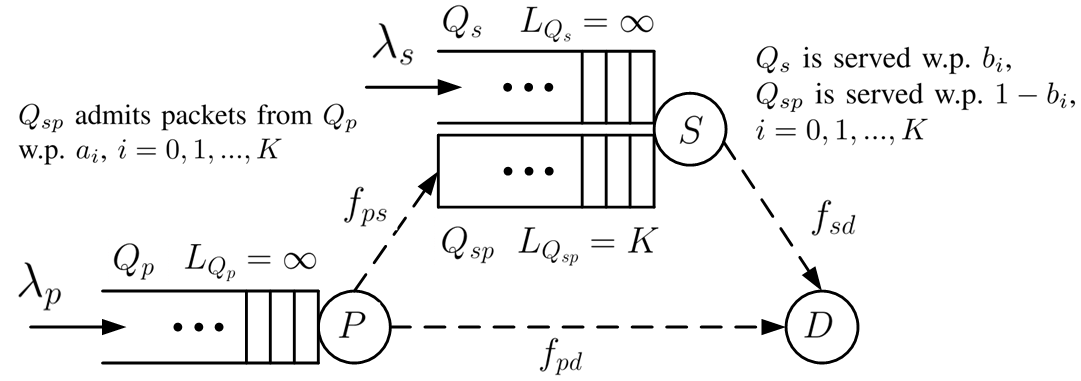}
\caption{The system model.}
\label{fig:SysModel}
\end{figure}
We consider the cooperative cognitive radio network shown in Fig.
\ref{fig:SysModel}. The network consists of a PU, an SU, and a
common destination. The PU is equipped with an infinite queue
($Q_p$) for its own packets. On the other hand, the SU is equipped
with an infinite queue ($Q_s$) for its own packets and a finite
queue ($Q_{sp}$) of size $K$ for the packets overheard, decoded and enqueued from the PU.
The system is time-slotted where the packet transmission from
either user to the destination takes precisely one time slot. The
external packet arrivals at $Q_p$ and $Q_s$ are modelled as
Bernoulli random processes with average arrival rates $\lambda_p$ and $\lambda_s$
packets per slots, respectively, where $0 \le \lambda_p, \lambda_s \le 1$.
The arrival processes of both users are assumed to
be independent from each other across time slots. Following the notation in 
\cite{Ashour_Journal}, The instantaneous length evolution of an arbitrary 
queue $i$ is defined as
\begin{equation}
Q_i^{t+1} = \left(Q_i^t - Y_i^t\right)^+ + X_i^t, \: i \in \{ p,
s, sp \}
\end{equation}
where $Q_i^t$ is the number of packets at the beginning of 
time slot $t$, and $X_i^t$ and $Y_i^t$ are the number
of arriving and departing packets in time slot $t$, respectively. 
The random variables, $X_i^t$ and $Y_i^t$, are binary taking 
values $0$ or $1$. Note that $(Z)^+ = \max(Z,0)$.

The quality of a wireless communication channel is deteriorated
due to the inherent impairments caused by signal
attenuation, shadowing, multipath fading, and additive noise. The
quality of the channel is abstracted in this work by the average channel
reception probability which is the probability that a transmitted
packet is correctly decoded. It is assumed that the SU 
has a perfect sensing capability and, hence, at
most one of the two users will transmit in every time slot.
Consequently, the only reason for packet loss is link
outage. Since we assume that the channel gain and 
noise processes are stationary, and the mobility of the nodes is
ignored in our analysis, the channel success probabilities are
deterministic and take values from the interval $[0,1]$. Let
$f_{pd}$, $f_{ps}$ and $f_{sd}$  denote the probability of
successful packet reception from the PU to the destination, from
the PU to the SU, and from the SU to the destination,
respectively. Throughout our analysis, we assume that $f_{pd} <
f_{sd}$, i.e., the SU has a better channel to the destination than
the PU. We will signify the prominence of such assumption to the
cooperation strategy later.

Acknowledgement packets (ACKs) are sent either by the destination
for successfully decoded packets from the PU or SU, or by the SU
for successfully-decoded overheard primary packets. 
We assume that ACKs are instantaneous as well as error-free. 
Additionally, ACKs can be heard by all the nodes in the system.

% wireless mulitple-access system under the framework of
% Intuition suggests that

%%%%%%%%%%%%%%%%%%%%%%%%%%%%%%%%%%%%%%%%%%%%%%%%%%%%%%%%%%%%%%%%%%%%%%%%%%%%%%%%%%%%%%%%%%%%%%%%%%%%%%%%%%%%%%%%%%%%%%%%%%%
\section{Cooperation with Finite Relaying Buffer at the SU}
The proposed system adopts the following cooperative policy at the
MAC layer. When the PU is backlogged, i.e., $Q_p$ is non-empty, it
transmits a packet. In such case, we have three possible
scenarios.  If the destination correctly decodes the PU packet, it
sends back an ACK that is heard by the PU and the packet is then
dropped from $Q_p$ and exits the system. On the other hand, if the
destination fails to decode the PU packet, but the SU correctly
decodes it, $Q_{sp}$ buffers the packet w.p. $a_i, \:
i=0,1,\ldots,K$. The packet admission probability in the system
depends on the number of packets in $Q_{sp}$. If the SU admits the
packet, it sends back an ACK so that the PU drops
the packet from its queue. Therefore, the SU is in charge of
transmitting this packet to the destination. Finally, if neither
the destination nor the SU successfully receives the PU packet, the
PU keeps this packet in its queue to be retransmitted in the next
time slot.

When the PU is idle, i.e., $Q_p$ is empty, the SU accesses the
channel. It transmits a packet either from $Q_{sp}$ w.p. $1-b_i,
\: i=0,1,\ldots,K$ or from $Q_s$ w.p. $b_i, \: i=0,1,\ldots,K$. The
queue selection probability in the system also depends on the
number of packets in $Q_{sp}$. If the destination correctly
decodes the SU packet, it sends back an ACK that is heard by the
SU. The packet is then dropped from either $Q_{sp}$ or $Q_s$ and
exits the system.

In view of the finite queue size of $Q_{sp}$, we should account
for the blocked packets when the relaying queue is full at a given
time slot, i.e., when it has $K$ packets. This case is handled by
forcing the system to not accept any relayed packets from the PU
when $Q_{sp}$ is full. In other words, we set $a_K=0$.
On the other hand, a time slot would be wasted if an empty
$Q_{sp}$ is selected for transmission while $Q_s$ still has
packets to transmit. In order to prevent such case, the system is
forced to select $Q_s$ for transmission whenever $Q_{sp}$ is
empty, In other words, $b_0=1$.

According to the aforementioned cooperation policy, the system 
at hand is non-work-conserving because there are cases
where the system might waste time slots in spite of having packets
to transmit. A typical case occurs when an empty $Q_s$ is
probabilistically selected for transmission while $Q_{sp}$ still
has packets to transmit. Thus, the system would waste such time
slots. Nonetheless, the non-work-conserving policy of our system
achieves the same stable throughput region of the work-conserving
policy characterized in \cite{R_CoopAcc} using an infinite relaying buffer. 
This result stems from optimally tuning the system degrees of freedom provided 
by the probabilistic packet admission and queue selection of 
the proposed cooperation policy with finite relaying buffer. 
The optimization problem and performance results supporting the 
above insights will follow in the next two sections.

%%%%%%%%%%%%%%%%%%%%%%%%%%%%%%%%%%%%%%%%%%%%%%%%%%%%%%%%%%%%%%%%%%%%%%%%%%%%%%%%%%%%%%%%%%%%%%%%%%%%%%%%%%%%%%%%%%%%%%%%%%%
\section{The Stable Throughput Region}
%In this section, we define the stable throughput region of the system.
\subsection{Queue Stability}
In a stable network of queues, every individual queue has to be stable.
Loynes' theorem establishes the condition for stability of an
infinite size queue \cite{L_QStab}. It asserts that if the queue arrival and
service processes are stationary, the queue is stable if and only
if the packet arrival rate $\lambda$ is strictly less than the
packet service rate $\mu$. It is worth mentioning that
$Q_{sp}$, with maximum size $K$, is stable for all positive values of the arrival and
service rates. The number of packets will never grow to infinity
since it is upper bounded by $K$. Let $\pi_i$, $i=0,1,\ldots,K$, be
defined as the probability that $Q_{sp}$ is in a given state $i$,
i.e., $Q_{sp}$ has $i$ packets, at a given time slot.

Note that a packet leaves $Q_p$ if it is either correctly decoded
by the destination or correctly decoded by the SU and 
admitted to the relaying queue. Therefore,
\begin{equation} \label{eq:mu_p}
\displaystyle \mu_p = f_{pd} + \left(1-f_{pd}\right) f_{ps}
\sum_{i=0}^K a_i\pi_i
\end{equation}
Also, a packet leaves $Q_s$ if the PU is idle, $Q_s$ is selected
for transmission, and the packet is correctly decoded by the
destination. Therefore,
\begin{equation} \label{eq:mu_s}
\displaystyle \mu_s = f_{sd} \left( 1 - \frac{\lambda_p}{\mu_p}
\right) \displaystyle\sum_{i=0}^K b_i\pi_i
\end{equation}
Accordingly, the stable throughput region of the system is characterized
as
\begin{multline} \label{eq:region}
\mathcal{R} = \biggl\{ (\lambda_p, \lambda_s) \enspace \Big| \enspace
\displaystyle\lambda_s < f_{sd} \left( 1 - \frac{\lambda_p}{\mu_p} \right) \displaystyle\sum_{i=0}^K b_i\pi_i,\\
\!\!\!\! \text{for } \displaystyle \lambda_p < f_{pd} + \left(1-f_{pd}\right)
f_{ps} \sum_{i=0}^K a_i\pi_i\biggr\}
\end{multline}
In pursuance of completely characterizing the stable throughput
region of the system, the steady-state distribution of $Q_{sp}$
should be calculated. In the next subsection, we conduct a
Discrete Time Markov Chain (DTMC) analysis for $Q_{sp}$.

\subsection{DTMC Analysis of $ Q_{sp}$ }
\begin{figure}
\centering
\includegraphics[width=1\linewidth]{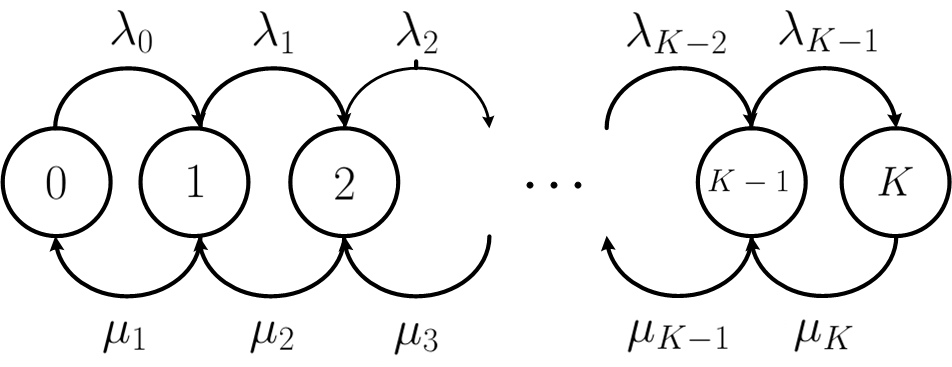}
\caption{Discrete Time Markov Chain (DTMC) model of the SU relaying queue, $Q_{sp}$.}
\label{fig:DTMC}
\end{figure}

Fig. \ref{fig:DTMC} depicts the state transition diagram of $Q_{sp}$ where
state $i$ denotes the number of packets in $Q_{sp}$. The
arrival rate $\lambda_i$ can be characterized as the conditional
probability that $Q_p$ is non-empty, the destination fails to
decode the PU packet, the SU correctly decodes that packet, and
$Q_{sp}$ admits the relayed packet, given that $Q_{sp}$ is at
state $i$. Thus,
\begin{equation} \label{lambda_i}
\displaystyle\lambda_i = \frac{\lambda_p}{\mu_p}
\left(1-f_{pd}\right) f_{ps} a_i, \: i = 0, 1,\ldots,K
\end{equation}
Similarly, $\mu_i$ can be characterized as the conditional
probability that $Q_p$ is idle, the SU selects $Q_{sp}$ to transmit
the enqueued PU packet, and the destination correctly decodes it, given that
$Q_{sp}$ is at state $i$. Thus,
\begin{equation} \label{mu_i}
\displaystyle\mu_i = \left(1-\frac{\lambda_p}{\mu_p}\right) 
\left(1-b_i\right) f_{sd}, \: i = 0, 1,\ldots,K
\end{equation}
It should be noted that $\lambda_K=0$ as $a_K=0$ and $\mu_0=0$ as
$b_0=1$.
%level of non working conserving and blocking

Using the balance equations, the steady-state probabilities can be expressed as
\begin{equation}\label{pi_j_1}
\displaystyle\pi_{j+1} = \frac{\lambda_j}{\mu_{j+1}}\pi_j
\end{equation}
where $j = 0,1,\ldots,K-1$. Substituting from (\ref{lambda_i}) and
(\ref{mu_i}) in (\ref{pi_j_1}), we get
\begin{equation} \label{pi_j_2}
\displaystyle\pi_{j+1}
=\frac{\lambda_p\left(1-f_{pd}\right)f_{ps}a_j}{\left(\mu_p-\lambda_p\right) \left(1-b_{j+1}\right) f_{sd}}\pi_j
\quad \text{for }j = 0,1,\ldots,K\!-\!1
\end{equation}
Applying the normalization condition,
\begin{equation}\label{normalize_probability}
    \displaystyle\sum_{i=0}^K \pi_i = 1
\end{equation}
along with (\ref{pi_j_2}), one can obtain the value of $\pi_0$ and, 
hence, completely characterize the steady-state distribution of
$Q_{sp}$.

\subsection{Stable throughput characterization problem}
Taking into consideration our foregoing discussion in the previous
two subsections, the problem of defining the stable throughput
region of the system can be formulated as maximizing the service
rate of the SU for a given arrival rate of the primary
traffic under stability constraints on all queues in the system,
i.e.,
\begin{eqnarray}
% \nonumber to remove numbering (before each equation)
  \max_{\mu_p, \{a_i, b_i, \pi_i\}} \!\!&&\!\! f_{sd} \left( 1 - \frac{\lambda_p}{\mu_p}
\right) \displaystyle\sum_{i=0}^K b_i\pi_i \nonumber \\
  \text{s. t. } \!\!&&\!\!   0 \leq a_i, b_i, \pi_i \leq 1 \qquad \text{for } i=0,1,\ldots, K  \nonumber \\
  \!\!&& \!\!  a_K=0,\: b_0=1   \nonumber\\
  \!\!&& \!\!  \mu_p = f_{pd} + \left(1-f_{pd}\right) f_{ps} \sum_{i=0}^K a_i\pi_i   \nonumber\\
  \!\!&& \!\!  \lambda_p < \mu_p \nonumber\\
  \!\!&& \!\!
   \eqref{pi_j_2},\eqref{normalize_probability}
  \label{max_1}
  \end{eqnarray}
Note that the optimization problem in (\ref{max_1}) is non-convex.
Nevertheless, we exploit the structure of the problem to transform
it into a linear program as follows.

First, we introduce the optimization variables  $x_i = a_i\pi_i$,
and $y_i = b_i\pi_i$, where $i \!=\! 0, 1,\ldots,K$. Therefore, we
can write \eqref{eq:mu_p} as
\begin{equation}
% \nonumber to remove numbering (before each equation)
 \mu_p \!=\!\ f_{pd} + \left(1-f_{pd}\right) f_{ps} \sum_{i=0}^K x_i  \label{constraint1}
  \end{equation}
Since $a_i, b_i, \pi_i \in [0,1]$, for $i=0,1,\ldots,K$, we have the
following constraints on $x_i$ and $y_i$,
\begin{equation}\label{range_constraints_x_y_1}
    0 \leq x_i, y_i \leq \pi_i \qquad\text{for } i=0, \ldots, K
\end{equation}
\begin{equation}\label{range_constraints_x_y_2}
	0 \leq \displaystyle\sum_{i=0}^K x_i \leq 1,\; 0 \leq \displaystyle\sum_{i=0}^K y_i \leq 1
\end{equation}
Furthermore, we can write the objective function of \eqref{max_1}
 and the constraint in \eqref{pi_j_2} as
\begin{eqnarray}
 \!\!\!\!\!\!&&\!\!\!\!\!\!\qquad\qquad\qquad\mu_s = f_{sd} \left( 1 -
\frac{\lambda_p}{\mu_p}
\right) \displaystyle\sum_{i=0}^K y_i \label{objective}\\
\!\!\!\!\!&&\!\!\!\!\!\!\qquad\left(\pi_{j\!+\!1} \!-\!
y_{j\!+\!1}\right)\! \left(\mu_p \!-\! \lambda_p\right) \! =\!
\frac{\lambda_p}{f_{sd}} \left(1 \!-\!
f_{pd}\right) f_{ps} x_j \nonumber\\
\!\!\!\!\!\!&&\!\!\!\!\!\!\qquad\qquad\qquad\qquad\qquad\qquad\qquad
\text{for } j\!=\!0,\ldots,K\!-\!1 \qquad \label{constraint2}
  \end{eqnarray}

%where we have used \eqref{constraint1}, and
%\begin{equation}\label{alpha}
%\alpha=\sum_{i=0}^K x_i
%\end{equation}

% after eliminating the variable $\mu_p$ and using the new variable $\alpha$

Next, we rewrite the optimization problem in (\ref{max_1}). The resulting problem is given by
\begin{eqnarray}
% \nonumber to remove numbering (before each equation)
  \max_{\mu_p, \{x_i, y_i, \pi_i\}} \!\!&&\!\! f_{sd} \left( 1 -
\frac{\lambda_p}{\mu_p} \right) \displaystyle\sum_{i=0}^K y_i \nonumber \\
  \text{s. t. } \!\!&&\!\!   0 \leq \pi_i \leq 1 \qquad \text{for } i=0,1,\ldots, K  \nonumber \\
  \!\!&& \!\!  x_K=0,\: y_0=\pi_0   \nonumber\\
  \!\!&& \!\!  \mu_p = f_{pd} + \left(1-f_{pd}\right) f_{ps} \sum_{i=0}^K x_i   \nonumber\\
  \!\!&& \!\!  \lambda_p < \mu_p \nonumber\\
  \!\!&& \!\!
   \eqref{normalize_probability},\eqref{range_constraints_x_y_1},\eqref{range_constraints_x_y_2},\eqref{constraint2}
  \label{max_2}
  \end{eqnarray}

The above optimization problem is still non-convex. However, at a given
value of $\mu_p$, the problem reduces into a linear program
in the variables $\{a_i, b_i, \pi_i\}$ which is given by
\begin{eqnarray}
% \nonumber to remove numbering (before each equation)
  \max_{\{x_i, y_i, \pi_i\}} \!\!&&\!\! f_{sd} \left( 1 -
\frac{\lambda_p}{\mu_p}
\right) \displaystyle\sum_{i=0}^K y_i \nonumber \\
  \text{s. t. } \!\!&&\!\!   0 \leq \pi_i \leq 1 \qquad \text{for } i=0,1,\ldots, K  \nonumber \\
  \!\!&& \!\!  x_K=0,\: y_0=\pi_0   \nonumber\\
  \!\!&& \!\!
   \eqref{normalize_probability},\eqref{range_constraints_x_y_1},\eqref{range_constraints_x_y_2},\eqref{constraint2}
  \label{max_3}
  \end{eqnarray}
It can be evidently shown from (\ref{eq:mu_p}) and (\ref{eq:region}) that
the feasible values of $\mu_p$ over which the linear program runs is
\begin{equation} \label{mu_p_ineq}
\max(\lambda_p,f_{pd}) \leq \mu_p \leq f_{pd}+\left(1-f_{pd}\right)f_{ps}
\end{equation}

To recapitulate briefly, we transformed the non-convex optimization problem
of finding the stable throughput region of the system
into a linear program via standard techniques. At a given throughput of the
PU, we run the resulting linear program in \eqref{max_3} under the feasible
values of $\mu_p$ given by \eqref{mu_p_ineq}. Our goal is to
identify the value of $\mu_p$ that corresponds to the maximum achievable
value of the objective function, i.e., the service rate of the SU,
while satisfying the system constraints.

%Consequently, we change the formulation of the optimization
%problem by fixing the value of $\mu_p$ and then running the
%optimization problem, for a given value of $\lambda_p$, at every
%permissible value of $\mu_p$.

%$\mu_p$ is a function of the optimization variables, namely
%$a_i$ and $\pi_i$, yet it can be evidently shown from (\ref{eq:mu_p}) and (\ref{eq:region}) that

%Theorem 1:  When  $f_{sd} \geq f_{pd}$, the optimal value of the
%objective function in \eqref{max_3} is monotonically increasing in
%the parameter $\alpha$.
%
%
%
%\emph{Proof:} See the Appendix for proof.

%One insightful conclusion drawn from observing the above problem
%is that the maximum achievable $\lambda_s$ at every $\lambda_p$ is
%always obtained when $\sum_{i=0}^K a_i\pi_i = 1$, regardless of
%the choice of $b_i, i = 1,...,K$. In other words, when $a_i=1$ for
%$i=0,1,...,K-1$. % (see \ref{app1}ppendix for proof). 
%This is the same conclusion reached when $Q_{sp}$ is of infinite size
%\cite{Ashour_Journal}. An intuitive explanation for this result is
%simple. From SU's point of view, it is always better off admitting
%PU's relayed packets, as long as the direct link of the SU to
%the destination is better than the one of the PU to the destination,
%in order to increase the likelihood that $Q_p$ is empty and hence
%the SU transmits its packets more often.

In the next section, we solve the aforementioned
linear program. The numerical results will reveal the underlying
trend of the optimal solution of the problem.

%efficiently

%%%%%%%%%%%%%%%%%%%%%%%%%%%%%%%%%%%%%%%%%%%%%%%%%%%%%%%%%%%%%%%%%%%%%%%%%%%%%%%%%%%%%%%%%%%%%%%%%%%%%%%%%%%%%%%%%%%%%%%%%%%
\section{Numerical Results}
In this section, we conduct a performance evaluation for the system under the proposed cooperative policy. The system parameters are the probabilities of no link outages between the different nodes that are selected to be $f_{pd}=0.3$, $f_{ps}=0.4$, $f_{sd}=0.8$. Moreover, a baseline comparison with the work-conserving scheme in which the SU fully cooperates with the PU via an infinite length queue \cite{R_CoopAcc} is presented. Furthermore, we used CVX, a package for solving convex optimization programs \cite{cvx}, \cite{gb08}, in order to solve the linear program in (\ref{max_3}).

\begin{figure}
\centering
\includegraphics[width=1\linewidth]{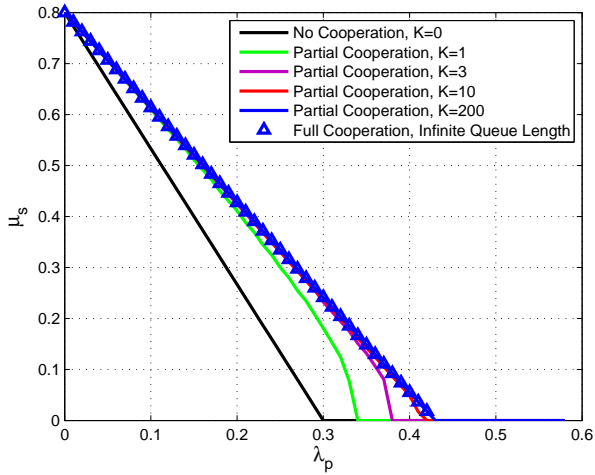}
\caption{The stable throughput regions of the system under different cooperative and non-cooperative schemes for different lengths of the relaying queue $Q_{sp}$ ($f_{pd}=0.3$, $f_{ps}=0.4$ and $f_{sd}=0.8$).}
\label{fig:throughout}
\end{figure}

Fig. \ref{fig:throughout} depicts the stable throughput region of 
the system for different lengths of the relaying queue $Q_{sp}$. 
The lower bound of the stable throughput region occurs when $K=0$.
This corresponds to 'No Cooperation' scenario whereby the SU doesn't 
cooperate with the PU in delivering its packets. 
On the other hand, the upper bound of the stable throughput region 
occurs when $K=200$, i.e., the queue length is relatively long with 
respect to the system parameters. This corresponds to 
'Full Cooperation, Infinite Queue Length' scenario \cite{R_CoopAcc}. 
Furthermore, As the queue length varies from $K=0$ to $K=200$, 
the stable throughput region increases as shown in Fig. \ref{fig:throughout}. 
In other words, on the assumption that $f_{sd} \geq f_{pd}$, 
when the level of cooperation in delivering the PU packets raises by 
increasing the length of the relaying queue $Q_{sp}$, the likelihood that 
$Q_p$ is empty increases and the SU can safely inject more packets 
into $Q_s$ without violating the system constraints. As a result, 
a corresponding increase in the throughput of the SU takes place. 
Moreover, it is obvious that the decrease in the stable throughput region, 
compared to the case of infinite relaying buffer size, is marginal 
even for relaying buffer sizes as small as 10 packets. 
In other words, the system doesn't lose much despite the limited relaying room.

%In Fig. \ref{fig:lemma}, the service rate of the PU is plotted
%against the service rate of the SU for the case of $K=200$ under
%the feasible values of $\lambda_p$. 
%It is evident that $\mu_s$ monotonically increases
%with $\mu_p$ irrespective of the value of $\lambda_p$. The 
%feasible values of $\mu_p$ is identified by (\ref{mu_p_ineq}).
%Correspondingly, the maximum achievable value of $\mu_s$ is
%obtained at the maximum achievable value of $\mu_p$ that is equal
%to $f_{pd}+\left(1-f_{pd}\right)f_{ps}$. According to
%(\ref{eq:mu_p}), it is required that $\sum_{i=0}^K a_i\pi_i = 1$
%in order to get this value and, thus, $a_i=1$ for $i=0,1,...,K-1$.
%This result matches the one reached by \cite{Ashour_Journal} for 
%the case of infinite relaying buffer size.

%This result matches our intuition as well as our rigorous
%mathematical proof presented in the the last part of the previous
%section.

Investigating the solution of our constrained optimization problem
of maximizing the SU throughput for a specific queue length $K$,
one can identify the underlying trend of the optimal solution of
the probability of selecting $Q_{sp}$, $1-b_i$, where $i=1,2,\ldots,K$.
As the PU injects more packets to the system, i.e., $\lambda_p$
increases, the probability of selecting $Q_{sp}$,
$1-b_i$, $i=1,2,\ldots,K$, increases. This is equivalent to
giving more priority to $Q_{sp}$ over $Q_s$ so as to accommodate the increase in
$\lambda_p$. This is a way to explains why $\lambda_s$ is
inversely proportional to $\lambda_p$. In addition to this partial cooperation solution, 
the full cooperation solution, $1-b_i=1$, $i=1,2,\ldots,K$,
also leads to the same maximum achievable stable throughput region. Our
solution of the optimization problem exploits the introduced
degree of freedom of the formulated problem that is represented in
the dependency of the queue selection probability on 
$Q_{sp}$ state at every time slot. Therefore, at
every value of $\lambda_p$, the solver assigns the minimum
$1-b_i$ that maximize the objective function
while satisfying all the system constraints. However, the full
cooperation solution assigns a fixed value for $1-b_i$
regardless of the value of $\lambda_p$.

%$\sum_{i=0}^K \left(1-b_i\right) \pi_i$
%Concerning the packet admission probability, the optimal solution requires admitting the primary packets w.p. 1 at every $\lambda_p$ in order to get the maximum $\mu_s$. This is equivalent to $\sum_{i=0}^K a_i\pi_i = 1$. The insight of this result has been presented in the last part of the previous section with a rigorous mathematical analysis to verify it.

%\begin{figure}
%\centering
%\includegraphics[width=1\linewidth]{LemmaFig}
%\caption{The relationship between $\mu_p$ and $\mu_s$ for the feasible values of $\lambda_p$ ($f_{pd}=0.3$, $f_{ps}=0.4$ and $f_{sd}=0.8$).}
%\label{fig:lemma}
%\end{figure}
%%%%%%%%%%%%%%%%%%%%%%%%%%%%%%%%%%%%%%%%%%%%%%%%%%%%%%%%%%%%%%%%%%%%%%%%%%%%%%%%%%%%%%%%%%%%%%%%%%%%%%%%%%%%%%%%%%%%%%%%%%%
\section{Conclusion}
We study the stable throughput region of a cooperative cognitive radio network when the relaying buffer at the SU has a finite capacity. We demonstrate a rigorous mathematical formulation for the problem. Although the formulation results in a constrained non-convex optimization problem, the problem is reformulated, via standard techniques, to be a linear program. Numerical results reveal the fact that cooperation of the SU in delivering the packets of the PU is always advantageous to both users in terms of expanding their stable throughput region. Furthermore, the numerical results show that the system does not lose much in terms of the stable throughput region despite the limited relaying capacity. Finally, solving the optimization problem yields a partial cooperation solution, in addition to the full cooperation solution, that maximizes the service rate of the SU due to the degrees of freedom introduced in the packet admission and queue selection probabilities. This is embodied in the dependency of such probabilities on the number of packets in the relaying queue. For future work, other performance metrics of the network can be studied in the light of the new dimensions added to the problem formulation.

\bibliographystyle{IEEEbib}
\linespread{1}
\bibliography{myRef}

\end{document}